\documentclass[%
 reprint,
amsmath,amssymb,
aps,
prc,
]{revtex4-2}

\usepackage{blindtext}
\usepackage{graphicx}
\usepackage{dcolumn}
\usepackage{bm}
\usepackage{hyperref}
\hypersetup
{
    colorlinks=true,
    urlcolor=black,
    linkcolor=black,
    citecolor=black,
}
\usepackage[mathlines]{lineno}

\begin{document}

\preprint{APS/123-QED}

\title{Anti-flow of Mesons in the High Baryon Density Region}

\author{Zuo-Wen Liu}
\affiliation{Key Laboratory of Quark and Lepton Physics (MOE) and Institute of Particle Physics, Central China Normal University, Wuhan 430079, China}

\author{Shusu Shi} 
\email[Corresponding author, ]{shiss@ccnu.edu.cn}
\affiliation{Key Laboratory of Quark and Lepton Physics (MOE) and Institute of Particle Physics, Central China Normal University, Wuhan 430079, China}

\date{\today}

\begin{abstract}
    
The E895 and STAR experiments demonstrate that the slopes of directed flow with respect to rapidity ($dv_1/dy|_{y=0}$) of mesons are negative in the low transverse momentum ($p_T$) region, $p_T < 0.8$ GeV/$c$, in Au + Au collisions at $\sqrt{s_{NN}} = 3.0 - 3.9$ GeV. Using the transport model JAM, we investigate the directed flow of $\pi^{\pm}$, $K^{\pm}$, and $K^0$ as functions of rapidity, $p_T$, and collision energy in Au + Au collisions at the same energies as those in the E895 and STAR experiments. We find that the JAM model can qualitatively reproduce the anti-flow of $K^0_S$ observed in the E895 experiment. The $v_1$ slopes of pions and kaons are analyzed as functions of the $p_T$ window, revealing a strong $p_T$ dependence of the $v_1$ slopes. Negative $v_1$ slopes are observed in the low $p_T$ region, $p_T < 0.8$ GeV/$c$, while positive slopes are shown in the higher $p_T$ region. We find that the shadowing effect from spectators is crucial in generating the anti-flow of mesons at low $p_T$ in the high baryon density region.

\end{abstract}

\maketitle

\section{Introduction}\label{sec.I}

The collective flow ($v_n$) is the event anisotropy of final state particles in heavy-ion collisions,
which can be defined by Fourier expansion of the particle distribution with respected to the reaction plane~\cite{Poskanzer:1998yz, Voloshin:2008dg}:
\begin{equation}
    E \frac{d^3 N}{d^3 p}=\frac{1}{2 \pi} \frac{d^2 N}{p_T d p_T d y}\left(1+\sum_{n=1}^{\infty} 2 v_n \cos \left[n\left(\phi-\Psi_r\right)\right]\right)
\label{eq:dNdPsi}
\end{equation}
where $\phi$, $\Psi_r$, and $v_n$ represent the azimuth of final state particle in the laboratory frame, the azimuth of the reaction plane, and the $n^{\rm th}$ harmonic coefficient of the Fourier series, respectively. 
The collective flow carries the information of the created medium in heavy-ion collisions~\cite{Hung:1994eq}, 
and is sensitive to the properties and the Equation of State (EoS) of the medium.
The first order coefficient, known as directed flow ($v_1$), signifies the sideward collective motion of particles along the direction of the impact parameter~\cite{Bilandzic:2010jr}.
In the context of heavy-ion collisions, $v_1$ offers valuable insights into the behavior of Quark-Gluon Plasma (QGP) and its transition to a hadronic phase. 
It acts as a sensitive probe for examining the properties of strongly interacting matter under extreme conditions in heavy-ion collisions~\cite{Gale:2012rq, Schenke:2010rr, Schenke:2010nt, Ryu:2021lnx, Ivanov:2020wak, Tsegelnik:2022eoz, Jiang:2023fad, Jiang:2023vxp, Karpenko:2023bok}. 
At the same time, it can reveal the interplay between initial compression and tilted expansion~\cite{Bozek:2010bi, Nara:2021fuu}, thereby playing an important role in exploring the Quantum Chromodynamics (QCD) phase diagram, especially in the high baryon density region with a large baryon chemical potential $\mu_B$~\cite{Bzdak:2019pkr, Chen:2024zwk, Luo:2020pef, Ivanov:2024gkn, Shen:2020jwv, Nara:2019qfd, Oliinychenko:2022uvy, Steinheimer:2022gqb, OmanaKuttan:2022aml, Li:2022cfd, Wu:2023rui, Yong:2023uct}. 
Moreover, $v_1$ measurements impose constraints on the EoS and identify its softest point in the QCD phase diagram~\cite{STAR:2014clz, STAR:2017okv, Stoecker:2004qu, Nara:2022kbb, Ivanov:2014ioa, Du:2022yok, Parfenov:2022brq, Mamaev:2023yhz, Kozhevnikova:2023mnw}. Thus, $v_1$ study in the high baryon density region is pivotal for enhancing our understanding of the QCD phase diagram and the nature of strongly interacting matter at extreme densities.

The sign of $v_1$ is commonly used to describe the deflection behavior of particles in heavy-ion collisions. A positive $v_1$ indicates that more freeze-out particles are emitted in the direction aligned with the spectators, while a negative $v_1$ indicates that more particles are emitted in the direction opposite to the spectators~\cite{Steinheimer:2014pfa}.
In high energy heavy-ion collisions, a negative $v_1$ at positive rapidity, leading to a negative slope at mid-rapidity ($dv_1/dy|_{y=0}$), known as anti-flow, can be attributed to the tilted expansion prevailing over the initial compression~\cite{Bozek:2010bi, Chatterjee:2017ahy, Bozek:2022svy, Jiang:2021ajc}. On the other hand, at lower collision energies ($\sqrt{s_{NN}} < 10$ GeV), the spectator shadowing effect becomes non-negligible, introducing another factor influencing the flow observable.
The shadowing effect generates negative elliptic flow, as observed in experiments such as Au + Au collisions at $\sqrt{s_{NN}}$ = 3.0 GeV from RHIC-STAR, where $\mu_B$ $\approx$ 720 MeV~\cite{STAR:2021yiu}. Regarding $v_1$, the anti-flow of $K^0_S$ was observed in Au + Au collisions at $\sqrt{s_{NN}}$ = 3.83 GeV by the E895 experiment at the Alternating Gradient Synchrotron (AGS) over two decades ago~\cite{E895:2000maf, E895:2000sor}. The kaon anti-flow is widely attributed to the repulsive kaon potential in the high baryon density region~\cite{Kaplan:1986yq, Brown:1991ig, Waas:1996fy, Schaffner:1995th, E895:2000sor, Li:1994vd, Pal:2000yc}.
It is also worth studying whether meson anti-flow is possibly caused by the spectator shadowing effect~\cite{Liu:1998yc}.

\begin{table}
    \centering
    \begin{tabular}{|c|c|c|c|c|c|}
        \hline
        $\sqrt{s_{NN}}$ (GeV) & $y$ &  $\beta$ ($c$)  &  $t_p$ (fm/$c$)  &  $t_m$ (fm/$c$)  & $t_p/t_m$ \\ \hline
        3.0 & 1.06 & 0.79 & 11.00 & 16.99 & 0.65 \\ \hline
        3.2 & 1.13 & 0.81 & 10.07 & 16.68 & 0.60 \\ \hline
        3.5 & 1.25 & 0.85 & 8.72  & 16.34 & 0.53 \\ \hline
        3.9 & 1.37 & 0.88 & 7.59  & 16.01 & 0.47 \\ \hline
    \end{tabular}
    \caption{Rapidity $y$, velocity $\beta$, the spectator passing time $t_p$, the hadron freeze-out mean time $t_m$ in JAM model, and their ratio $t_p/t_m$ from Au + Au collisions at $\sqrt{s_{NN}}$ = 3.0, 3.2, 3.5, and 3.9 GeV. Note that $t_m$ is calculated from all hadrons over $|\eta| < 1 $.}
    \label{tab:tableI}
\end{table}

The passing time of spectators ($\sim 2R/\gamma\beta$) closely matches the average time of particle freeze-out~\cite{Bialas:1988mn, Liu:1998yc, Lin:2017lcj}, where $R$ is the radius of nuclei, $\gamma$ is the Lorentz contraction factor, and $\beta$ is the velocity of the spectators as a fraction of the speed of light. It leads to freeze-out particles being shadowed by the spectators.
Table~\ref{tab:tableI} lists the spectator passing times $t_p$ in the center-of-mass frame for Au + Au collisions at $\sqrt{s_{NN}}$ = 3.0, 3.2, 3.5, and 3.9 GeV. The average time of particle freeze-out $t_m$ is estimated using the JET AA Microscopic Transport Model (JAM)~\cite{Nara:1999dz}. The ratio of $t_p$ to $t_m$ indicates that the spectator passing time is about half of the average time of particle freeze-out in Au + Au collisions at $\sqrt{s_{NN}}$ = 3.0 - 3.9 GeV. Furthermore, this ratio increases as the collision energy decreases, suggesting that the spectator shadowing effect becomes more significant at lower collision energies.
Using the JAM model, we investigate the shadowing effect on the $v_1$ of pions and kaons in Au + Au collisions at $\sqrt{s_{NN}}$ = 3.0 - 3.9 GeV, corresponding to a $\mu_B$ coverage from 630 to 720 MeV. We find that anti-flow is not only observed for $K^0$ but also for charged pions and kaons in the low $p_T$ region.

The rest of this paper is organized as follows. We introduce the JAM model in Section~\ref{sec.II}. Then, in Section~\ref{sec.III}, we present a comparison of $K_S^0$ anti-flow from the E895 experiment to JAM calculations, along with a detailed study of $v_1$ slopes of pions and kaons with the JAM model at the same collision energies as the STAR experiment~\cite{Liu:2023tqz}. To study the shadowing effect from spectators, we compare the results with and without spectators from JAM model. Finally, we provide a summary in Section~\ref{sec.IV}.

\section{JAM Model}\label{sec.II}

\begin{figure}[hbt!]
\centering
\includegraphics[width=1.0\linewidth]{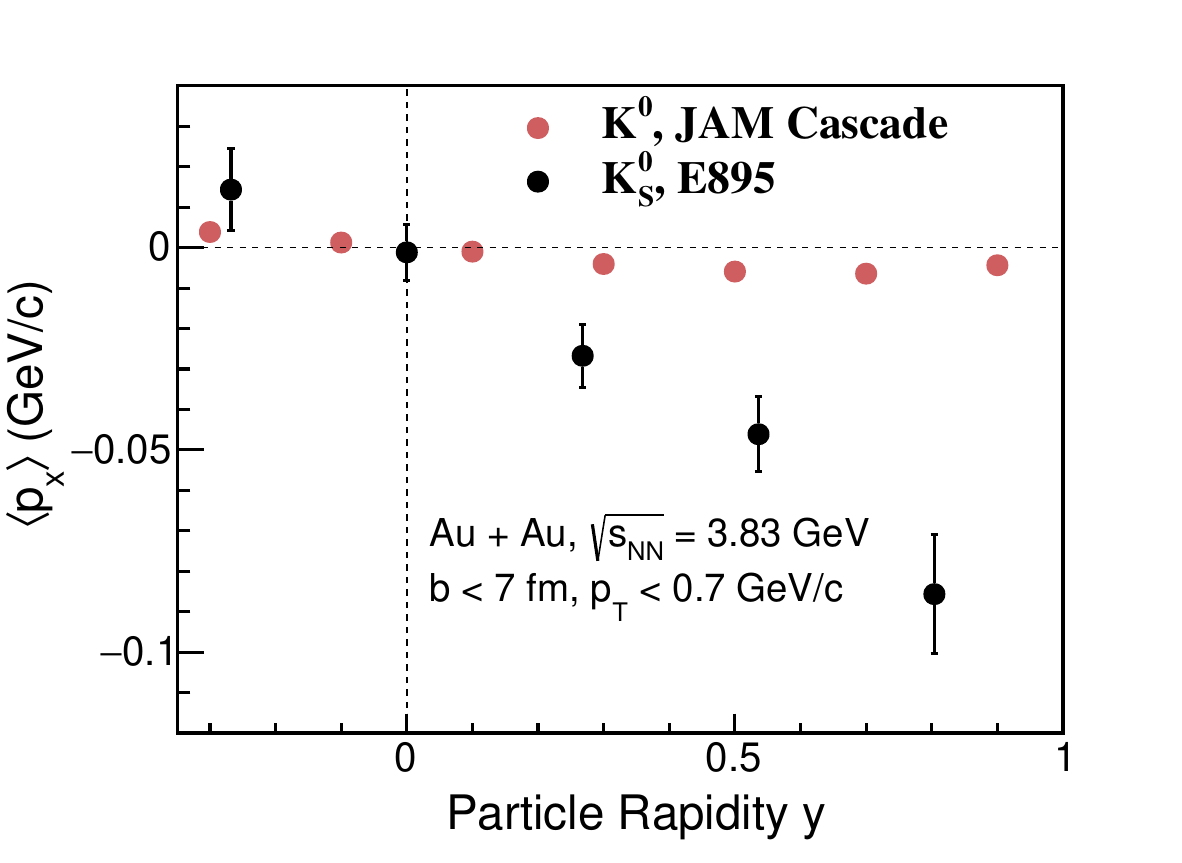}
\caption{The directed transverse flow $\langle p_x \rangle$ of $K^{0}$ from JAM calculations (red circles), and  that of $K^{0}_{S}$ from E895 experiment (black circles) as a function of rapidity in Au + Au collisions at $\sqrt{s_{NN}}$ = 3.83 GeV.
Note that the normalized rapidity from E895 data has been converted to the center-of-mass frame to enable a direct comparison.}
\label{fig:comp_v1}
\end{figure}

\begin{figure}[hbt!]
\centering
\includegraphics[width=1.0\linewidth]{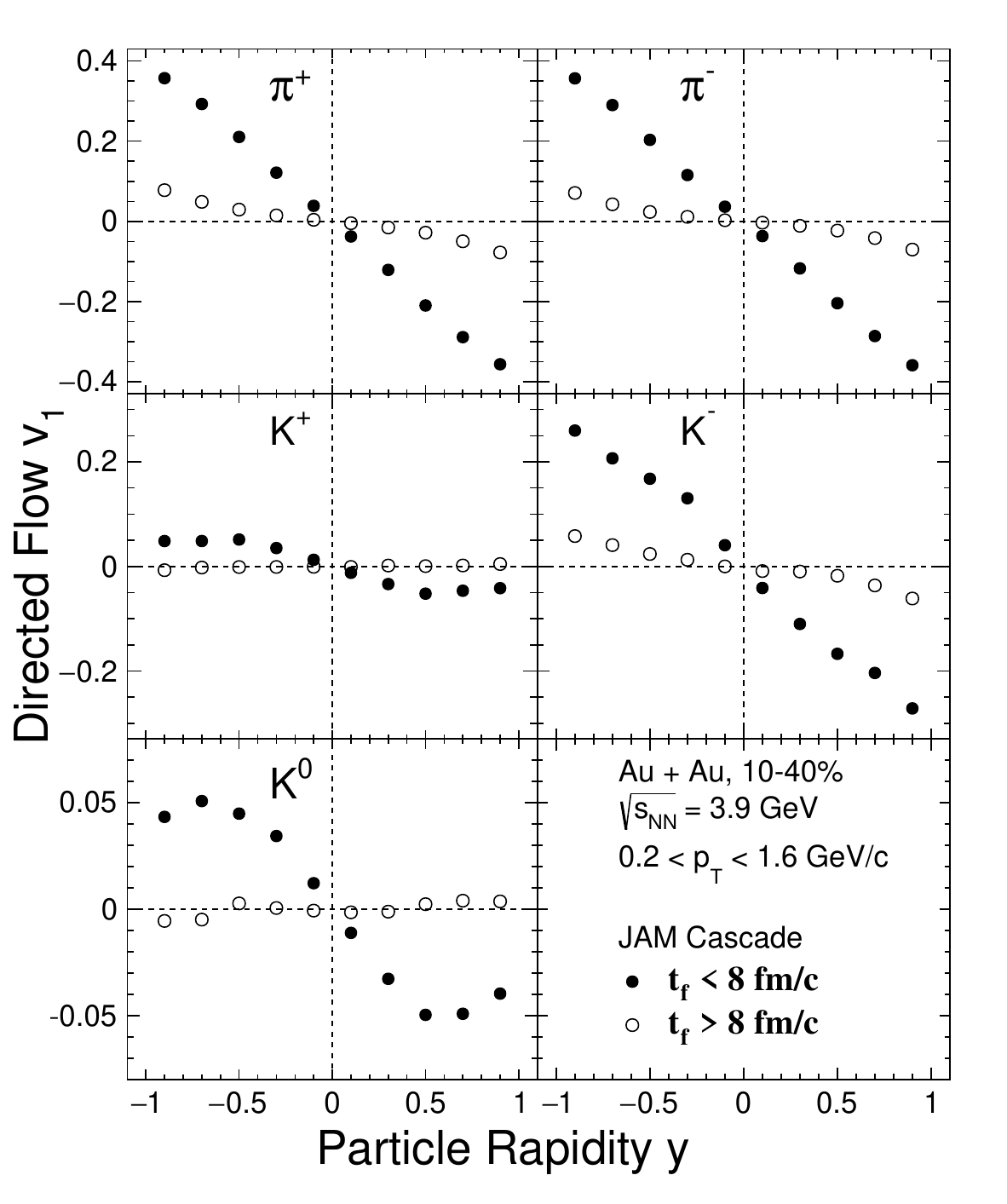}
\caption{The directed flow as a function of rapidity in 10-40\% most central Au + Au collisions at $\sqrt{s_{NN}}$ = 3.9 GeV from JAM cascade mode,
where black solid and open circles denote the particles which freeze-out time is less and more than 8 fm/$c$, respectively.
The left and right panels depict results of particles ($\pi^{+}$, $K^{+}$, and $K^{0}$) and the corresponding anti-particles ($\pi^{-}$ and $K^{-}$), respectively.}
\label{fig:v1VSy_Tf}
\end{figure}

\begin{figure}[hbt!]
\centering
\includegraphics[width=1.0\linewidth]{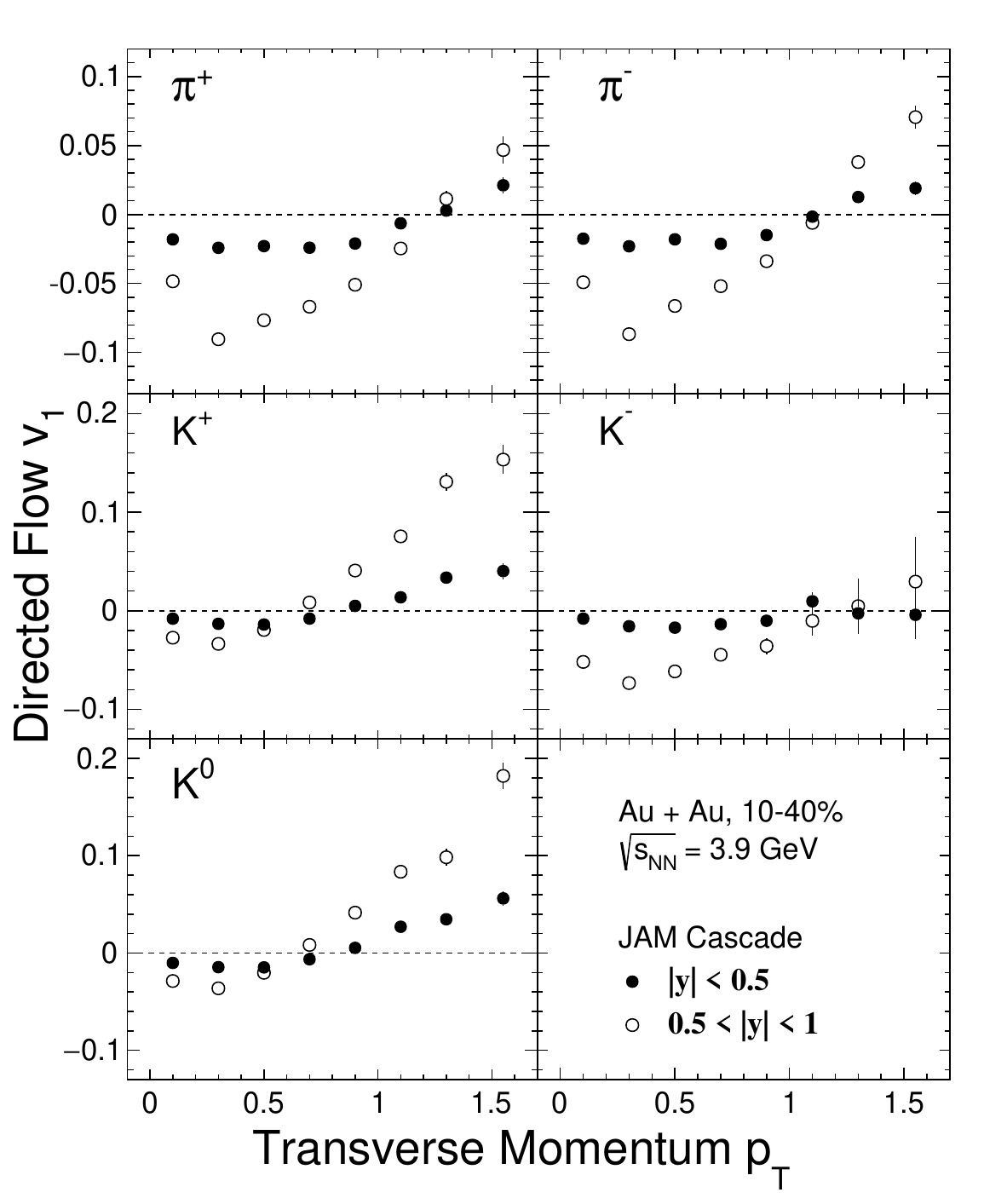}
\caption{The directed flow as a function of $p_T$ in 10-40\% most central Au + Au collisions at $\sqrt{s_{NN}}$ = 3.9 GeV from JAM cascade mode,
where black solid circles are for particles in the mid-rapidity ($|y| < 0.5$) and open circles for forward-rapidity ($0.5 < |y| < 1$).
The left and right panels depict results of particles ($\pi^{+}$, $K^{+}$, and $K^{0}$) and the corresponding anti-particles ($\pi^{-}$ and $K^{-}$), respectively.}
\label{fig:v1VSpT}
\end{figure}

The JAM model is a hadronic transport model capable of simulating the evolution of relativistic heavy-ion collisions~\cite{Nara:1999dz}. In this model, particle production includes resonance excitation, string production, and their decay contributions, similar to other transport models such as RQMD, AMPT, and PHSD~\cite{Sorge:1995dp, Lin:2004en, Nayak:2019vtn, Cassing:2009vt}.
Two different approaches are used to describe the effect of the EoS in JAM. One is the cascade method based on the modified two-body scattering~\cite{Hirano:2012yy}, while the other involves the nuclear mean-field method implemented with the relativistic quantum molecular dynamics approach~\cite{Sorge:1989dy}. 

It has been demonstrated that the baryonic mean-field potential plays a significant role in the high baryon density region.
JAM model calculations cooperating with baryonic mean-field have successfully described the $v_1$ and $v_2$ of baryons in Au + Au collisions at $\sqrt{s_{NN}}$ = 3.0 GeV from RHIC-STAR~\cite{STAR:2021yiu, Lan:2022rrc}.
However, the mean-field mode is not suitable for the $v_1$ of mesons in the high baryon density region, probably due to the implementation of the strong repulsive baryonic potential~\cite{Nara:2020ztb}. 
The cascade mode describes the $v_1$ of pions and kaons better than the baryonic mean-field mode at $\sqrt{s_{NN}} < 4.5$ GeV, as illustrated in Refs.~\cite{STAR:2021yiu, Parfenov:2022brq, MPD:2021kbc, Zhang:2018wlk, Nara:2021fuu, Nara:2020ztb}.
Therefore, we will focus on the JAM model in cascade mode for the following meson $v_1$ study.
Additionally, JAM provides an option to retain or remove spectators during the nuclear matter evolution in heavy-ion collisions. We take advantage of this feature to investigate the non-trivial shadowing effect caused by spectators in the high baryon density region.

In this work, we use version 2.1 of the JAM model to generate Monte Carlo event samples for Au + Au collisions at $\sqrt{s_{NN}}$ = 3.0, 3.2, 3.5, and 3.9 GeV. We employ the default cascade mode and the cascade mode without spectators to study the anti-flow of mesons in heavy-ion collisions within the high baryon density region. Event centrality is defined by the reference multiplicity, which counts the total number of charged pions, charged kaons, protons, and anti-protons within the pseudo-rapidity region of $|\eta| < 0.5$, which is same as STAR experimental analysis~\cite{Liu:2023tqz}.
The azimuth of the reaction plane is zero in the model, thus the directed flow can be calculated by $v_1 = \langle \cos(\phi) \rangle$. 
In experiments, as the reaction plane is unknown, the Event Plane method is used for the $v_1$ measurements~\cite{Poskanzer:1998yz}. 
Using the JAM model, we compare the results from the Event Plane method, following the STAR experiment~\cite{STAR:2021yiu}, to those of $\langle \cos(\phi) \rangle$. 
The results from the two methods exhibit good consistency in the model. 
The statistical error is smaller in the case of $\langle \cos(\phi) \rangle$ as it does not require resolution correction due to the event plane estimate. We will present the results of $\langle \cos(\phi) \rangle$ in the following.

\section{Results and Discussions}\label{sec.III}
We compare the measurements of directed transverse flow $\langle p_x \rangle$ as a function of rapidity from the E895 experiment with JAM calculations. The experimental measurement of $\langle p_x \rangle$ extracted from Au + Au collisions at $\sqrt{s_{NN}}$ = 3.83 GeV by the E895 Collaboration~\cite{E895:2000maf,E895:2000sor} is shown as black circles in Fig.~\ref{fig:comp_v1}. To facilitate a direct comparison, the normalized rapidity from E895 data has been converted to the rapidity in the center-of-mass frame~\cite{E895:2000maf,E895:2000sor}. The JAM calculations with cascade mode are denoted by red circles, and they are applied the same impact parameter range and transverse momentum cut as the data. Note that $K^0$ in JAM includes both $K^0_S$ and $K^0_L$ since the model cannot separate $K^0_S$ and $K^0_L$. The model can reproduce the anti-flow of $K^0_S$ in mid-rapidity but underestimates the $v_1$ strength. The discrepancy between the model and the data might be due to the lack of a kaon potential in the JAM cascade mode. Since the model can reproduce the negative $v_1$ slope of $K^0_S$ in mid-rapidity as observed in the experimental data, we will further investigate the impact of the shadowing effect with the JAM model, which is important for the meson $v_1$ in the high baryon density region.

\begin{figure}[hbt!]
\centering
\includegraphics[width=1.0\linewidth]{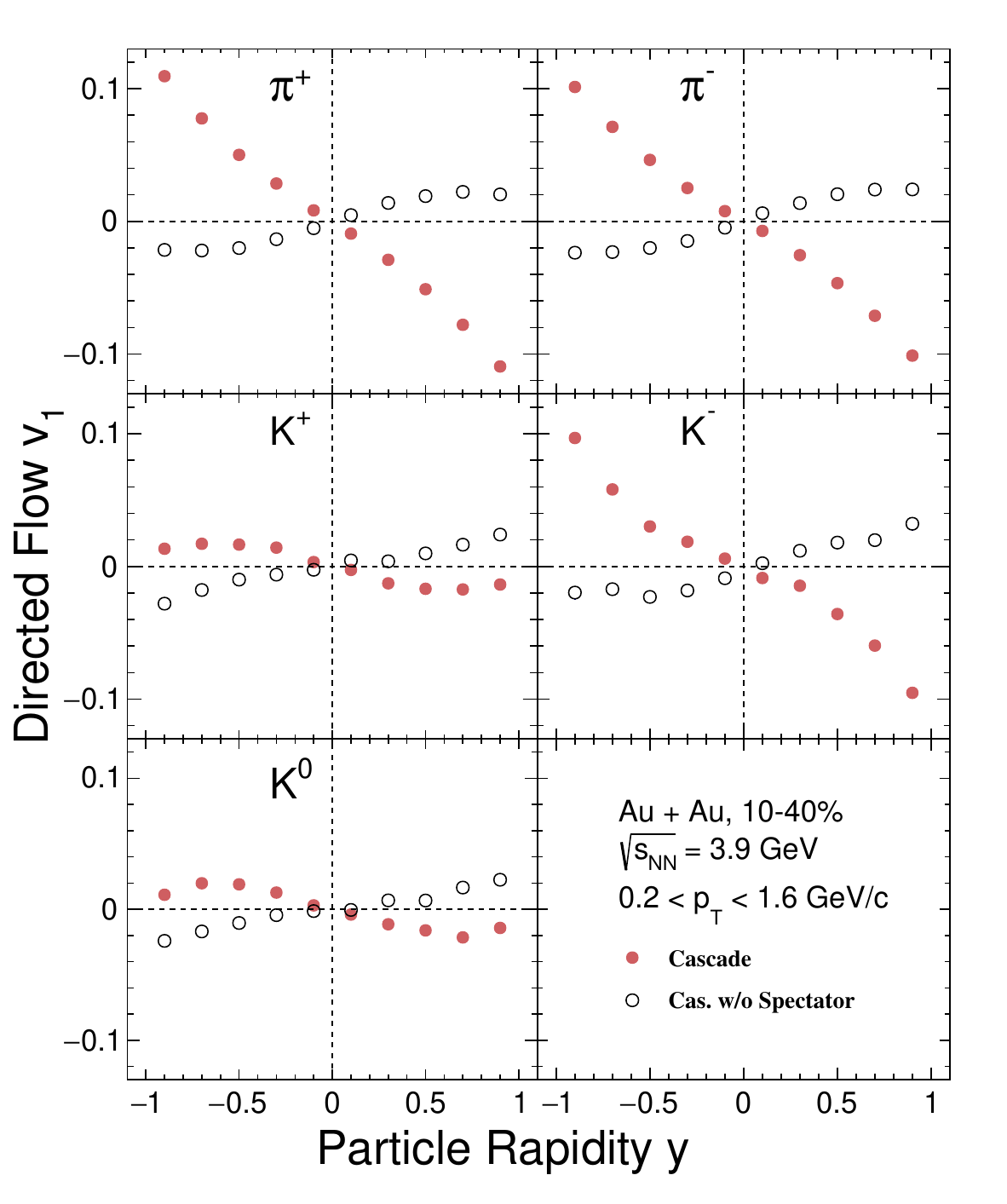}
\caption{The directed flow as a function of rapidity in 10-40\% most central Au + Au collisions at $\sqrt{s_{NN}}$ = 3.9 GeV
from JAM in default cascade mode (red solid circles) and in cascade mode without spectator (open circles).
The left and right panels depict results of particles ($\pi^{+}$, $K^{+}$, and $K^{0}$) and the corresponding anti-particles ($\pi^{-}$ and $K^{-}$), respectively.
}
\label{fig:v1VSy}
\end{figure}

\begin{figure}[hbt!]
\centering
\includegraphics[width=1.0\linewidth]{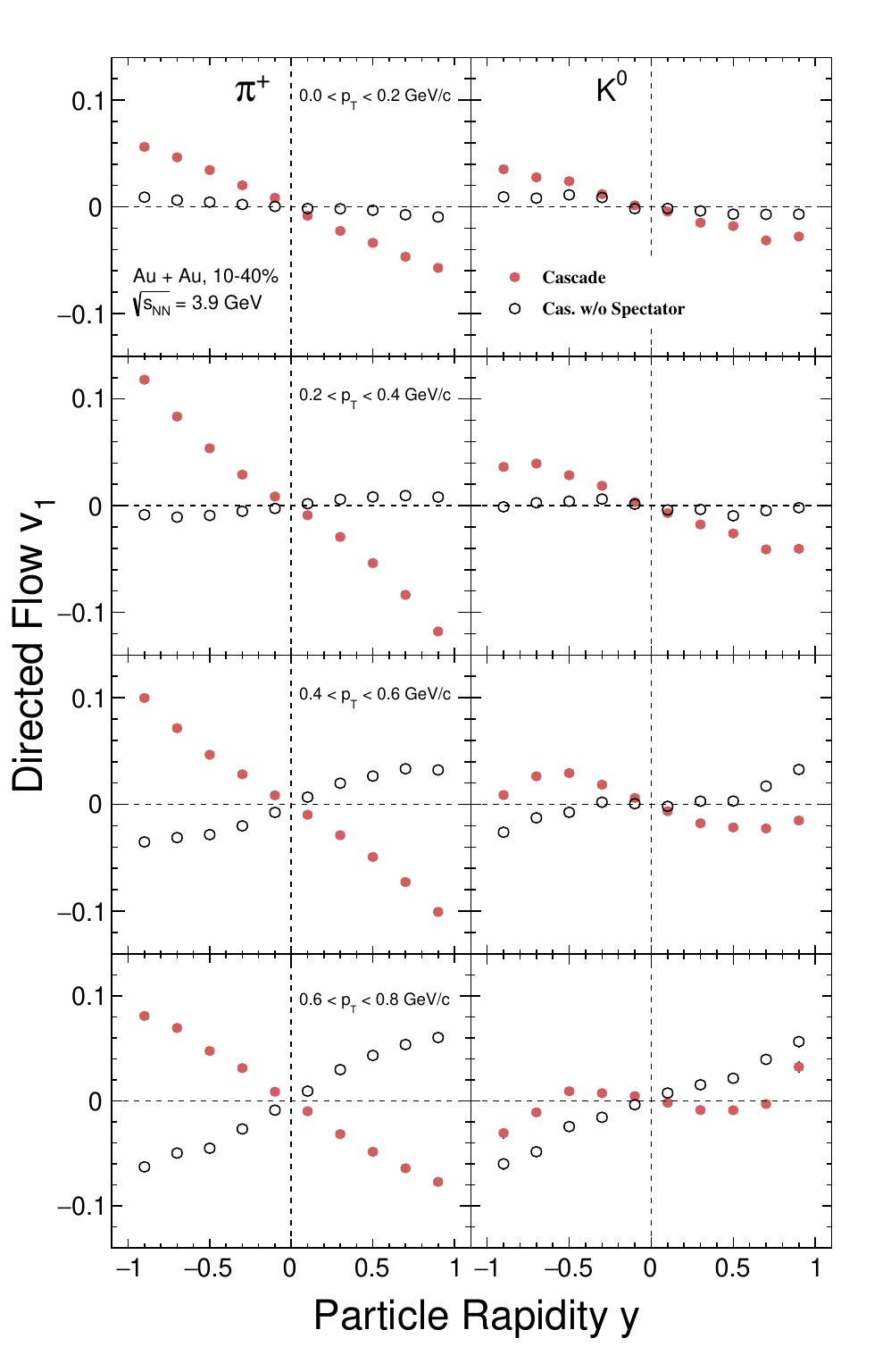}
\caption{The directed flow as a function of rapidity in different $p_T$ windows for $\pi^{+}$ (left panels) and $K^{0}$ (right panels)
in 10-40\% most central Au + Au collisions at $\sqrt{s_{NN}}$ = 3.9 GeV
from JAM cascade mode (red solid circles) and cascade mode without spectator (open circles).}
\label{fig:v1y_pT}
\end{figure}

\begin{figure*}[hbt!]
\centering
\includegraphics[width=0.9\linewidth]{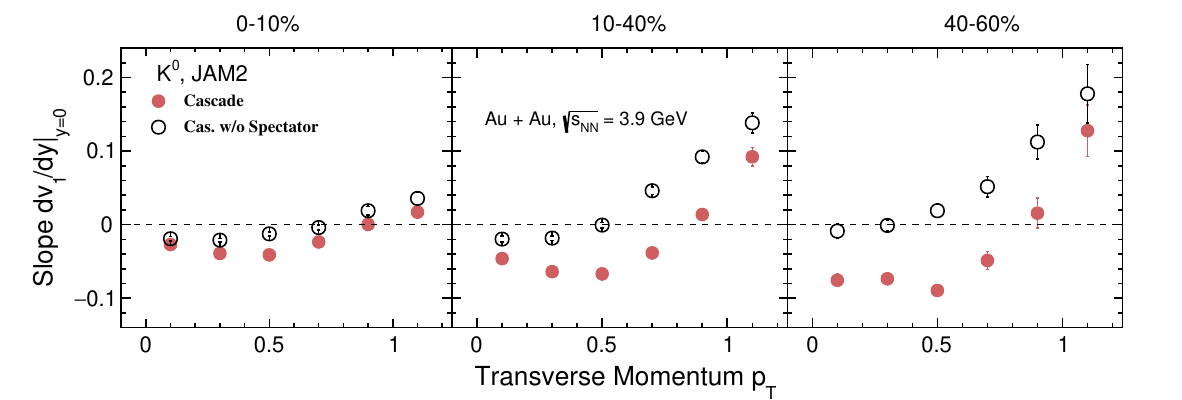}
\caption{The $v_1$ slopes at mid-rapidity ($dv_1/dy|_{y=0}$) as a function of $p_T$ window for $K^{0}$
in 0-10\%, 10-40\% and 40-60\% most central Au + Au collisions at $\sqrt{s_{NN}}$ = 3.9 GeV
form JAM cascade mode (red solid circles) and cascade mode without spectator (open circles).
The $dv_1/dy|_{y=0}$ is characterized by the linear term in a fit of the function $v_1(y) = ay + by^3$, the fit range is over -1 $< y <$ 0.}
\label{fig:v1Slope_pT}
\end{figure*}

\begin{figure}[hbt!]
\centering
\includegraphics[width=1.0\linewidth]{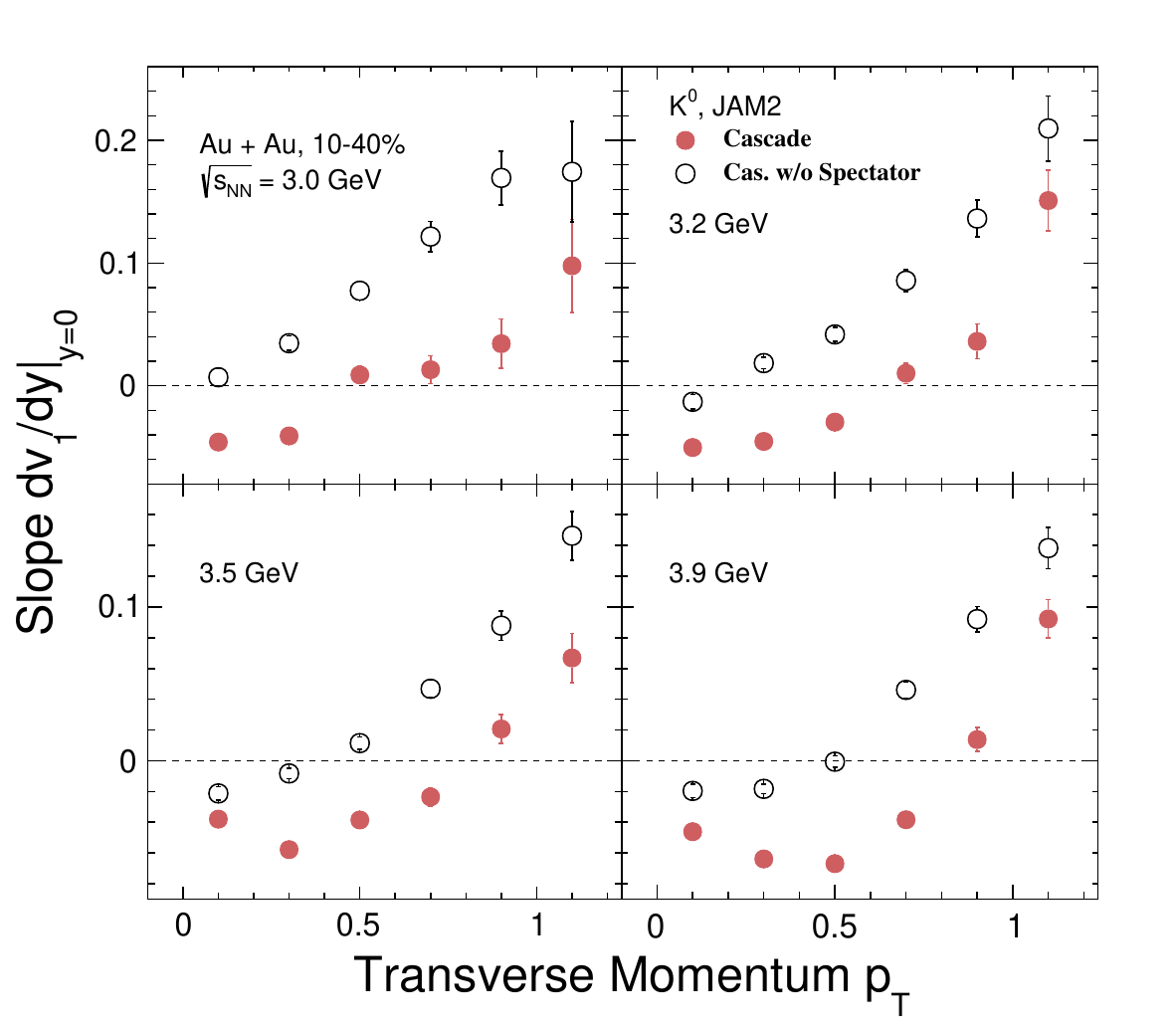}
\caption{The $v_1$ slopes at mid-rapidity ($dv_1/dy|_{y=0}$) as a function of $p_T$ window for $K^{0}$
in 10-40\% most central Au + Au collisions at $\sqrt{s_{NN}}$ = 3.0, 3.2, 3.5, and 3.9 GeV
form JAM cascade mode (red solid circles) and cascade mode without spectator (open circles).}
\label{fig:v1Slope_sNN}
\end{figure}

First, we explore the time evolution of directed flow. As shown in Table~\ref{tab:tableI}, the spectator passing time is about 8 fm/$c$ in Au + Au collisions at $\sqrt{s_{NN}}$ = 3.9 GeV. It is anticipated that particles with a freeze-out time of less than 8 fm/$c$ would be shadowed by the passing spectators, potentially influencing the development of directed flow.
Fig.~\ref{fig:v1VSy_Tf} depicts the freeze-out time dependence of $v_1(y)$ for pions ($\pi^{+}$ and $\pi^{-}$) and kaons ($K^{+}$, $K^{-}$, and $K^{0}$) in the 10-40\% most central Au + Au collisions at $\sqrt{s_{NN}}$ = 3.9 GeV from JAM calculations. The solid circles represent the particles that freeze out prior to the passage of the spectator, and the open circles represent those that freeze out after the spectator has passed.
Distinct negative $v_1$ slopes around mid-rapidity are observed for pions and kaons with a freeze-out time of less than 8 fm/$c$, and the $v_1$ slopes approach zero for pions and kaons with freeze-out times exceeding 8 fm/$c$. This implies that particles freezing out earlier experience a significant shadowing effect from spectators, leading to the observed anti-flow. As time progresses and the shadowing effect weakens, the $v_1$ slopes approach zero for particles with later freeze-out times.

In addition to the spectator shadowing effect being evident in the time evolution, it is expected that particles in the forward rapidity region exhibit greater sensitivity to the shadowing effect compared to those in the mid-rapidity region, as the former are in closer proximity to the spectator~\cite{Zhang:2018wlk, Nara:2022kbb}. Note that the spectator rapidity is 1.37 at $\sqrt{s_{NN}} = 3.9$ GeV, as shown in Table~\ref{tab:tableI}.
Fig.~\ref{fig:v1VSpT} illustrates the transverse momentum dependence of $v_1$ from mid-rapidity (solid circles) and forward rapidity (open circles) in 10-40\% most central Au + Au collisions at $\sqrt{s_{NN}} = 3.9$ GeV, obtained from JAM calculations. There is a clear negative $v_1$ at low $p_T$ for pions ($\pi^{+}$ and $\pi^{-}$) and kaons ($K^{+}$, $K^{-}$, and $K^{0}$) at mid-rapidity ($|y| < 0.5$). The value of $v_1$ in the forward rapidity region ($0.5 < |y| < 1$), close to the spectator rapidity, shows more negativity. This indicates that a well-defined anti-flow is formed at low $p_T$.
Furthermore, The $v_1$ values for pions exhibit more negativity at low $p_T$ compared to those for kaons, which could be attributed to the significantly larger scattering cross-section of pions relative to kaons ($\sigma_{K_0+p} \approx 10$ mb and $\sigma_{\pi+p} \approx 100$ mb)~\cite{E895:2000sor}.

The time evolution and $p_T$ dependence of $v_1$ suggest a negative contribution to $v_1$ of mesons arising from the spectator shadowing effect. Further study is carried out by removing the spectators in the JAM model.
We investigate the $v_1$ of pions and kaons in an overall transverse momentum ($p_T$) window (0.2 $< p_T <$ 1.6 GeV/$c$), which is the same as the acceptance range of the STAR experiment~\cite{Liu:2023tqz}.
In Fig.~\ref{fig:v1VSy}, the rapidity dependence of directed flow for particles ($\pi^{+}$, $K^{+}$, and $K^{0}$) and their corresponding anti-particles ($\pi^{-}$ and $K^{-}$) in 10-40\% Au + Au collisions at $\sqrt{s_{NN}}$ = 3.9 GeV is shown for JAM calculations in default cascade mode and cascade mode without spectators. It is observed that the JAM in cascade mode predicts negative mid-rapidity $v_1$ slopes for pions and kaons. On the other hand, the JAM in cascade mode without spectators shows positive $v_1$ slopes, in contrast to the one with spectator interactions. This implies that the shadowing effect from spectators leads to anti-flow of mesons in heavy-ion collisions in the high baryon density region. 
We studied $dv_1/dy|_{y=0}$ of pions ($\pi^{+}$ and $\pi^{-}$) and kaons ($K^{+}$, $K^{-}$, and $K^{0}$) in the JAM, and found that there is no strong charge dependence for the formation of anti-flow. To facilitate the following discussion, we take $\pi^{+}$ and $K^{0}$ as examples.

Then, we investigate the $v_1$ distribution in the narrow transverse momentum windows.
 The rapidity dependence of $\pi^{+}$ and $K^{0}$ $v_1$ within different $p_T$ windows in 10-40\% Au + Au collisions at $\sqrt{s_{NN}}$ = 3.9 GeV is presented in Fig.~\ref{fig:v1y_pT}. JAM in cascade mode depicts negative $v_1$ slopes at mid-rapidity for $\pi^{+}$ and $K^{0}$ $v_1$ at $p_T < 0.8$ GeV/$c$. The magnitude of $v_1$ slopes from the cascade mode without spectators decreases compared to those with spectators in the top two panels of Fig.~\ref{fig:v1y_pT} (with a $p_T$ window of $0 < p_T < 0.2$ GeV/$c$). As the $p_T$ increases, $v_1$ slopes of $\pi^{+}$ and $K^{0}$ from cascade mode without spectators turn positive rapidly, and clear positive $v_1$ slopes can be observed when $0.6 < p_T < 0.8$ GeV/$c$ as shown in the bottom two panels of Fig.~\ref{fig:v1y_pT}. Meanwhile, the $v_1$ slopes of $\pi^{+}$ and $K^{0}$ from default cascade mode remain negative. The $p_T$ dependence of $v_1(y)$ and the discrepancy between cascade mode with and without spectators indicate that, the shadowing effect from spectators is one of the important reasons for the anti-flow of $\pi^{+}$ and $K^{0}$.

Besides the $p_T$ window, the shadowing effect is also sensitive to the collision centrality. The number of spectator nucleons is larger in more peripheral collisions, thus suggesting that the shadowing effect should be stronger. Fig.~\ref{fig:v1Slope_pT} shows the $p_T$ dependence of $v_1$ slope at mid-rapidity for $K^{0}$ across three event centralities (0-10\%, 10-40\%, and 40-60\%) in Au + Au collisions at $\sqrt{s_{NN}}$ = 3.9 GeV. It is clearly observed that anti-flow for $K^{0}$ occurs at low $p_T$ in three centralities ($p_T < 0.8$ GeV/$c$ for 0-10\%, $p_T < 0.6$ GeV/$c$ for 10-40\%, and $p_T < 0.4$ GeV/$c$ for 40-60\%) in JAM cascade mode both with and without spectator interactions. The JAM with spectators pushes the $v_1$ slopes of $K^{0}$ to be more negative compared to the same model without spectators. Moreover, the magnitude of the $v_1$ slope in peripheral collisions (40-60\%) is greater than that in central collisions (0-10\%) at the low $p_T$ region, due to a stronger shadowing effect caused by the increased number of spectators in peripheral collisions. The strong centrality dependence of the $v_1$ slope confirms that the observed anti-flow of $K^{0}$ at low $p_T$ could be attributed to the shadowing effect from spectators.

The STAR experiment recently presented preliminary results on identified particle $v_1$ in Au + Au collisions at $\sqrt{s_{NN}} = 3.0$, 3.2, 3.5, and 3.9 GeV, where anti-flow of mesons was observed at low $p_T$ in these preliminary results~\cite{Liu:2023tqz}. Fig.~\ref{fig:v1Slope_sNN} shows $v_1$ slopes at mid-rapidity of $K^{0}$ as a function of $p_T$ in Au + Au collisions with 10-40\% centrality at $\sqrt{s_{NN}}$ = 3.0, 3.2, 3.5, and 3.9 GeV from JAM model calculations. The anti-flow of $K^{0}$ at low $p_T$ ($p_T < 0.8$ GeV/$c$) is captured at these four collision energies by the JAM in cascade mode. However, for the JAM cascade mode without spectators, the $v_1$ slopes are closer to zero from negative direction compared to those with spectators. In particular, the anti-flow of $K^{0}$ cannot be observed at all in the shown $p_T$ range at $\sqrt{s_{NN}} = 3.0$ GeV in JAM without spectators, and the difference in $v_1$ slope is more evident between JAM with and without spectators. This is probably because the strongest shadowing effect experienced by $K^0$, characterized by the largest $t_p/t_m$ ratio in Tab.~\ref{tab:tableI} among the four collision energies, is not accounted for in JAM without spectators. The comparison between JAM mode with and without spectators implies that the shadowing effect from spectators plays an important role in the kaon anti-flow at low $p_T$ in non-central heavy-ion collisions within the high baryon density region.

\section{Summary}\label{sec.IV}
In summary, we present calculations of the directed flow of $\pi^{\pm}$, $K^{\pm}$, and $K^0$ in Au + Au collisions at $\sqrt{s_{NN}}$ = 3.0, 3.2, 3.5, and 3.9 GeV using the hadronic transport model JAM. The JAM cascade mode can qualitatively reproduce the $K^0_S$ anti-flow observed by the E895 experiment in Au + Au collisions at $\sqrt{s_{NN}}$ = 3.83 GeV. Pions and kaons exhibit negative mid-rapidity $v_1$ slopes in the low $p_T$ region, $p_T < 0.8$ GeV/$c$, while the slopes become positive in the higher $p_T$ region. 
The time evolution of $v_1$ reveals that the pions and kaons with earlier freeze-out time exhibit distinct negative $v_1$ slopes around mid-rapidity, contrasting with a flat $v_1$ distribution from those with later freeze-out time.
And the $v_1$ values in the forward-rapidity are more negative compared to those in the mid-rapidity.
Furthermore, the centrality dependence of the $v_1$ slope shows that slopes are more negative in more peripheral collisions, which aligns with the expected spectator shadowing effect.
To investigate the shadowing effect, we analyze the $v_1$ slopes of mesons using the JAM model in cascade mode with and without spectator. The findings indicate that the shadowing effect from spectators turns positive $v_1$ slopes into negative ones in the low $p_T$ region at $\sqrt{s_{NN}}$ = 3.0 - 3.9 GeV.

It suggests that in heavy-ion collisions in the high baryon density region, the shadowing effect, arising from the presence of spectators, influences the motion of particles, leading to an anti-flow phenomenon of mesons where these particles move opposite to the normal flow direction. The kaon potential, often considered as a significant factor in understanding kaon anti-flow, may not be the sole or primary source. The shadowing effect could also contribute significantly. It also implies that the dynamics of high-density nuclear collisions are complex, with multiple factors influencing the collective flow observable. The shadowing effect from spectators plays an important role in understanding the observed phenomena.

\begin{acknowledgments}
We are grateful for discussions with Drs. Sooraj Radhakrishnan and Nu Xu.
This work is supported in part by the National Key Research and Development Program of China under Contract No. 2022YFA1604900; the National Natural Science Foundation of China (NSFC) under contract No. 12175084.
\end{acknowledgments}

\bibliographystyle{apsrev4-2} 
\bibliography{ref}


\end{document}